\documentclass{article}

\usepackage[utf8]{inputenc} 
\usepackage[T1]{fontenc}    
\usepackage{hyperref}       
\usepackage{url}            
\usepackage{booktabs}       
\usepackage{amsfonts}       
\usepackage{nicefrac}       
\usepackage{microtype}      
\usepackage{cleveref}       
\usepackage{graphicx}
\usepackage{xcolor}         

\usepackage[
style = nature,
sorting = none,
maxcitenames = 6, 
url = false,
block = none,
]{biblatex}
\hypersetup{
	colorlinks,
	linkcolor={black},
	citecolor={blue!50!black},
	urlcolor={blue!80!black}
}

\bibliography{Bibliography.bib}

\AtEveryBibitem{%
\clearfield{language}
\clearlist{language}
\clearfield{month}
\clearfield{series}
\clearfield{note}
\clearfield{issn}
}

\usepackage{doi}
\usepackage{setspace}
\title{Controlling 3D spin textures by manipulating sign and amplitude of interlayer DMI with electrical current}
\date{}
\author{}

\begin{document}
\maketitle

\noindent
Fabian Kammerbauer\textsuperscript{1},
Won-Young Choi\textsuperscript{1,2},
Frank Freimuth\textsuperscript{1,3},
Kyujoon Lee\textsuperscript{4},
Robert Frömter\textsuperscript{1},
Dong-Soo Han\textsuperscript{2},
Henk J. M. Swagten\textsuperscript{5},
Yuriy Mokrousov\textsuperscript{1,3}
and Mathias Kläui\textsuperscript{1}

\bigskip\noindent
\textbf{1}  Institute of Physics, Johannes Gutenberg-Universität Mainz, Mainz, Germany
\\
\textbf{2}  Center for Spintronics, Korea Institute of Science and Technology, Seoul, Republic of Korea
\\
\textbf{3}  Peter Grünberg Institut and Institute for Advanced Simulation, \linebreak Forschungszentrum Jülich and JARA, Jülich, Germany
\\
\textbf{4}  Division of display and semiconductor physics, Korea University, Sejong, \linebreak Republic of Korea
\\
\textbf{5} Department of Applied Physics, Institute for Photonic Integration, Eindhoven University of Technology, Eindhoven, The Netherlands
\bigskip

\newpage
\begin{abstract}

The recently discovered interlayer Dzyaloshinskii-Moriya interaction (IL-DMI) in multilayers with perpendicular magnetic anisotropy favors the canting of spins in the in-plane direction and could thus enable new exciting spin textures such as Hopfions in continuous multilayer films.
A key requirement is to control the IL-DMI and so 
in this study, the influence of an electric current on the IL-DMI is
investigated by out-of-plane hysteresis loops of the anomalous Hall effect
under applied in-plane magnetic fields. The direction of the in-plane field
is varied to obtain a full azimuthal dependence, which allows us to quantify
the effect on the IL-DMI. We observe a shift in the azimuthal dependence
of the IL-DMI with increasing current, which can be understood from the additional
in-plane symmetry breaking introduced by the current flow. Using an empirical model
of two superimposed cosine functions we demonstrate the presence of
a current-induced term that linearly increases the IL-DMI in the direction
of current flow. With this, a new easily accessible possibility to manipulate
3D spin textures by current is realized. As most spintronic devices
employ spin-transfer or spin-orbit torques to manipulate spin textures,
the foundation to implement current-induced IL-DMI into thin-film devices is broadly available.

\end{abstract}
\newpage
\section{Introduction}

The burgeoning field of 3D magnetism has introduced ways of bending 
and forming magnetic structures to ones will and produced magnetic 
nanoparticles~\cite{lopez-ortega_applications_2015}, magnetic 
tubes~\cite{streubel_retrieving_2015} and 
more~\cite{fischer_launching_2020, fernandez-pacheco_three-dimensional_2017}.
Each with the distinct feature of manipulating the material itself towards 
its use in the third dimension. Another approach towards 3D magnetism are 3D
topological textures within the magnetic system 
itself~\cite{fischer_launching_2020}. These new 3D spin structures, such as 
hopfions~\cite{kent_creation_2021}, offer advanced options for future 
spintronic 
applications~\cite{fischer_launching_2020, fernandez-pacheco_three-dimensional_2017}.
Compared to 2D spin structures, e.g. skyrmions stabilized by the 
Dzyaloshinskii-Moriya interaction (DMI)~\cite{fert_skyrmions_2013},
it is more challenging to stabilize twisted spin configurations in the third dimension. 
The recently experimentally observed magnetic Hopfion uses a combination 
of strong DMI and the confinement in a nanodisk~\cite{kent_creation_2021}.\\
Another route towards Hopfions is offered by the 
antisymmetric interlayer exchange interaction, which has been termed as 
interlayer DMI 
(IL-DMI)~\cite{han_long-range_2019, fernandez-pacheco_symmetry-breaking_2019, vedmedenko_interlayer_2019, avci_chiral_2021}. 
It favors spin canting between layers in the lateral direction 
and originates from an in-plane symmetry breaking. This allows for a highly
sought-after tuning mechanism for 3D magnetic textures, 
which could stabilize such structures in magnetic multilayers without confinement.
The absence of confinement opens the possibility to move or aggregate these structures.
While it is interesting to study the internal dynamics 
under the application of fields or currents~\cite{liu_three-dimensional_2020}
a possibility to directly control the IL-DMI strength can
additionally provide direct access to internal transformations 
between different 3D magnetic textures. This would allow for writing and deleting operations, 
which are necessary when employing the different magnetic textures as bits.
Conventionally, the strength of both the DMI and the IL-DMI is set during sample growth. 
For DMI post-growth changes to its magnitude have been shown, 
while for the IL-DMI these are elusive so far. 
One possible option to influence the DMI is strain. 
It has been shown that strain can enhance the DMI~\cite{deger_strain-enhanced_2020, filianina_thesis}
and be used to tune the diameter of skyrmions~\cite{shen_strain_2021}.
A technologically more easily accessible option is 
applying an electrical field, which is typically used to 
introduce spin-transfer torques and spin-orbit torques in many 
systems~\cite{ hirsch_spin_1999,miron_perpendicular_2011, emori_current-driven_2013, manchon_current-induced_2019}
and has been recently shown to directly influence the strength of the conventional DMI
~\cite{karnad_modification_2018, kato_current-induced_2019, freimuth_dynamical_2020}.
The benefit of this approach is the direct tuning parameter 
allowing one to locally change the DMI and therefore the spin structures 
can be dynamically controlled. \\
So far, no tuning of the IL-DMI after growth has been reported but it is a key step towards using this effect.
The goal of this study is to 
investigate possible effects on the IL-DMI induced by electrical currents.
This will be a more direct and locally accessible tuning parameter for the IL-DMI
rather than the global in-plane symmetry breaking
or varying the spin-orbit-coupling by a different material choice,
which cannot be changed after growth.
Following this thought, this study investigates the influence of an electrical
current on the IL-DMI by employing asymmetric hysteresis loop measurements,
a technique we introduced earlier to demonstrate the presence of the 
IL-DMI~\cite{han_long-range_2019}.
We use symmetry arguments to explain that a chiral Néel-type
state requires a gradient along the direction $(\textbf{M}_1\times \textbf{M}_2)\times \textbf{R}$,
where $\textbf{R}$ is the distance vector between the magnetizations $\textbf{M}_1$
and $\textbf{M}_2$ of the
two layers. When this gradient is provided by an applied electric field,
several mechanisms for IL-DMI become possible. One mechanism is the spin Hall effect,
which extends the spin-current picture of equilibrium
DMI~\cite{kikuchi_dzyaloshinskii-moriya_2016, freimuth_relation_2017}
into the nonequilibrium regime.

\section{Sample preparation}
\label{sec:Sample_preparation}
The investigated sample is a synthetic antiferromagnet (SAF) of
Ta(4)/Pt(4)/ Co(0.6)/Pt(0.7)/Ru(0.8)/Pt(0.5)/Co(1)/Pt(4) structure on Si/SiO$_2$
(layer thicknesses in nanometers). The two ferromagnetic (FM) layers,
FM1 at the bottom and FM2 at the top, are antiferromagnetically
coupled across the Ru layer via the Ruderman-Kittel-Kasuya-Yosida (RKKY)
interaction. The in-plane symmetry breaking
necessary to give rise to the 
IL-DMI~\cite{fernandez-pacheco_symmetry-breaking_2019, han_long-range_2019},
is introduced randomly during the sputtering process for this sample. 
The film was patterned into Hall bars by electron-beam lithography
to allow for higher current densities compared to a continuous-film measurement. 

\section{Method}
\begin{figure}[htb!]
    \centering
    \includegraphics[width = \linewidth]{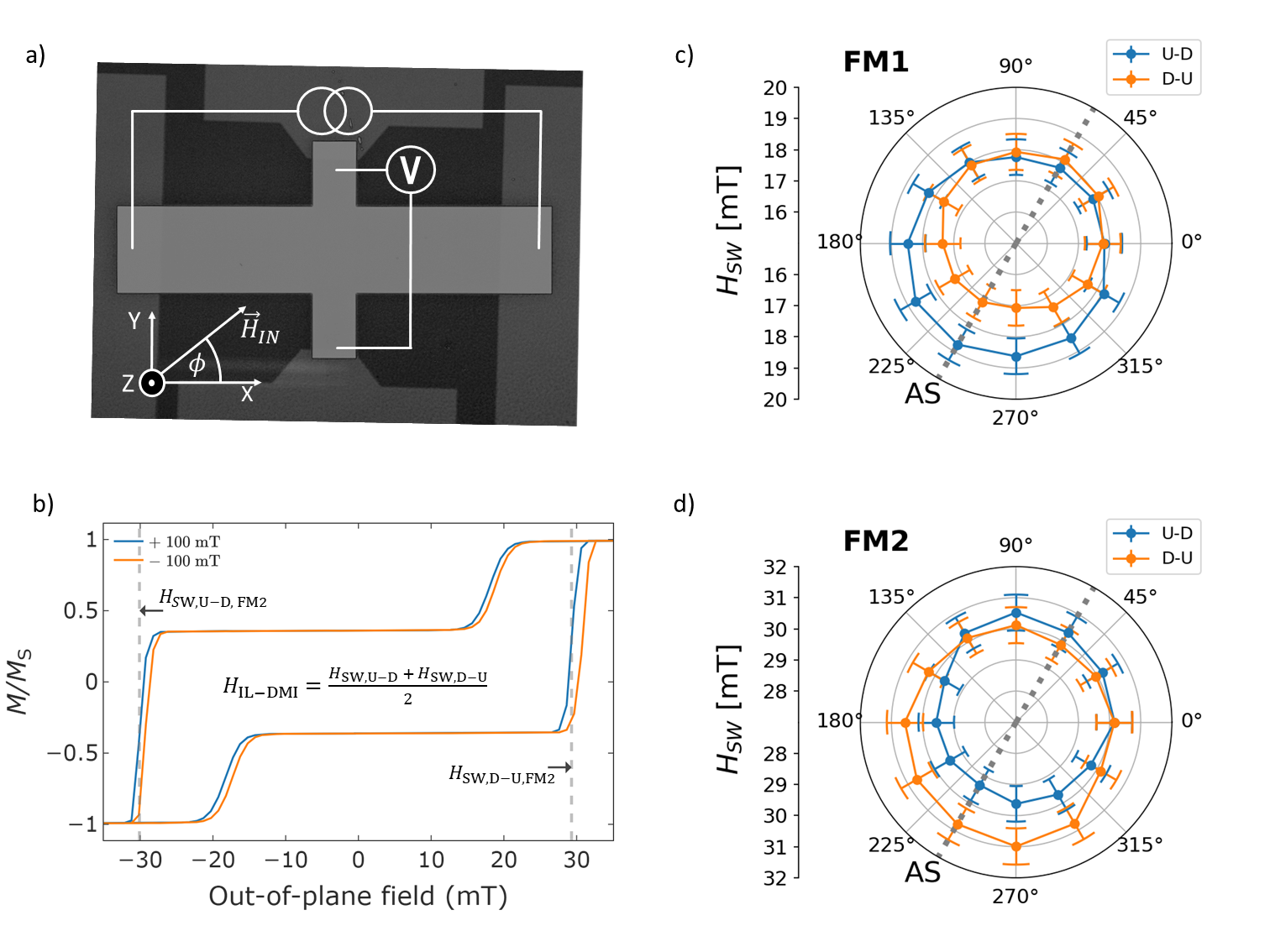}
    \caption{a) Hall-bar structure with indicated connections.
    For the anomalous Hall effect measurements a dc current along
    $+x$ direction is applied while sweeping the out-of-plane field
    in $z$-direction. Additionally, an in-plane field  $\textbf{H}_{\rm{IN}}$ of 100 mT
    is applied along an angle $\phi$, with $\phi$ varying
    from 0° to 330° in 30° steps. b) Magnetic hysteresis loops
    measured by anomalous Hall effect. Blue and orange curves
    display the hysteresis under the influence of positive and negative
    in-plane field of 100 mT, respectively, which was aligned along the
    antisymmetric axis. The difference between the two switching
    fields can be defined as a bias field $H_{\rm{IL-DMI}}$ originating
    from the IL-DMI. c) and d) display the dependence of 
    the switching fields on the in-plane field angle $\phi$ for
    FM1 and FM2 respectively measured at the lowest current density
    of $0.6\,\times 10^9 $A/m$^2$. The antisymmetric axis (AS) is indicated
    by a grey dotted line.}
    \label{fig:Figure_1_sample}
\end{figure}

To investigate the IL-DMI, the method devised by 
Han et al.~\cite{han_long-range_2019} was employed.
For the case of a sample with IL-DMI and perpendicular magnetic anisotropy (PMA)
the magnetization switching by an out-of-plane (OOP) 
field sweep will be supported or hindered in the presence
of an in-plane bias field depending on its direction and the chirality of the 
DMI~\cite{han_asymmetric_2016}. 
Similarly, in SAFs with PMA and IL-DMI, the magnetization
switching is supported or hindered under application of
an in-plane field depending on its azimuthal direction
with respect to the in-plane symmetry breaking axis and
the chirality introduced by the IL-DMI.
Hysteresis loops obtained by sweeping the OOP field were measured
by the anomalous Hall effect (AHE) in the patterned Hall bars while
applying an in-plane field of 100 mT along a varied direction $\phi$.
The structure of the connections for the AHE setup are shown in 
\cref{fig:Figure_1_sample}a). Two resulting hysteresis loops are presented in
\cref{fig:Figure_1_sample}b). The switching fields are derived from the
hysteresis curve by fitting error functions to the data.
The experimental uncertainty for all obtained switching-field values
is primarily given by the step size of 0.5 mT for the hysteresis loops.
The obtained switching fields can be determined separately
for the two magnetic layers as the different thicknesses yield different signal height.
In addition the two distinct sweep directions up-to-down (U-D)
and down-to-up (D-U), with up and down positive and negative field
in z-direction, are considered separately.
By varying $\phi$ in steps of 30° a full azimuthal dependence is obtained.
This allows us to find the direction of the axis of the antisymmetric
behaviour, which is the direction where the asymmetry between U-D and D-U
switching fields is the largest, as shown
in \cref{fig:Figure_1_sample}c) and d).
There the antisymmetric axis points approximately along 240°.
The presence of an antisymmetric axis indicates the presence of IL-DMI in this sample.

\section{Results}
\label{sec:results}

\begin{figure}[htb!]
    \centering
    \includegraphics[width=\linewidth]{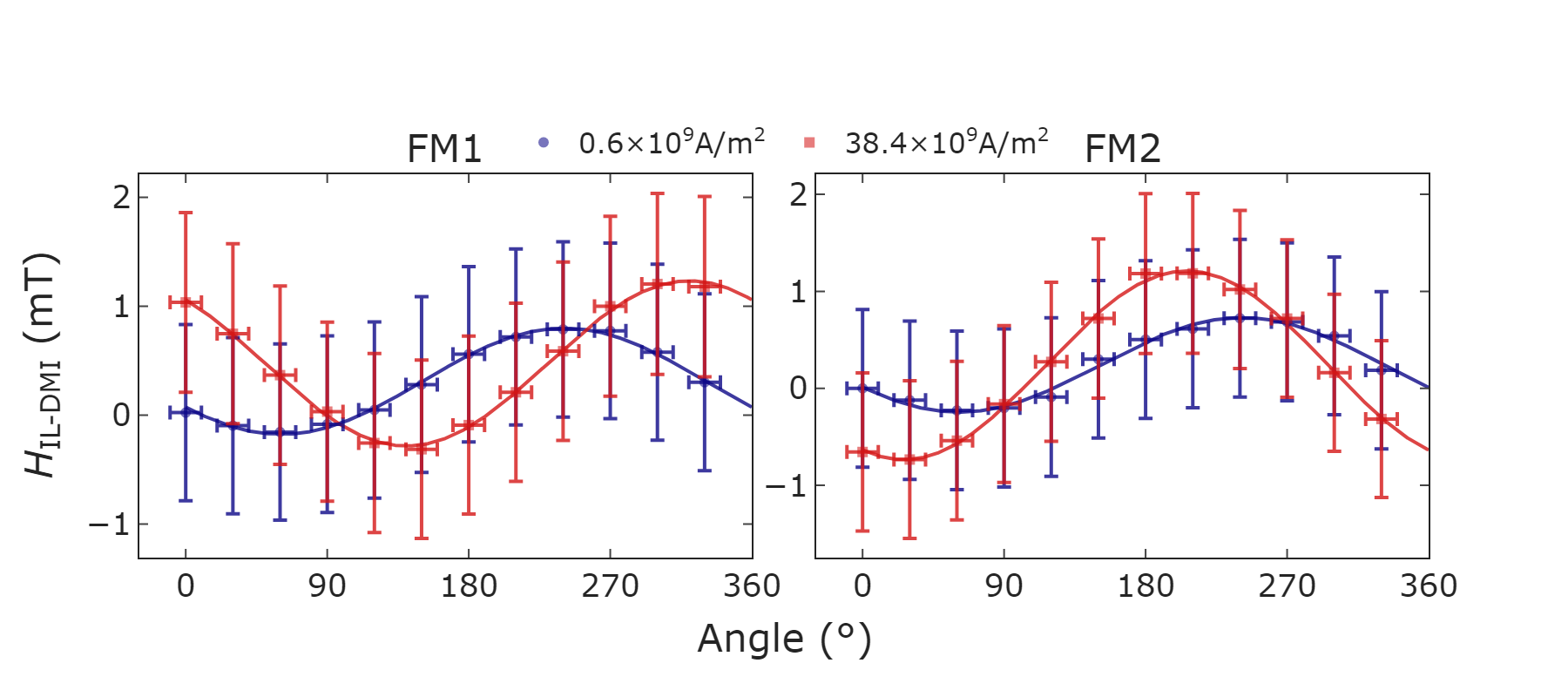}
    \caption{Azimuthal dependence of IL-DMI for the lowest current density
    at $0.6\,\times 10^9  $A/m$^2$ and the highest current density at $38.4\,\times 10^9 $A/m$^2$,
    for FM1 on the left  and FM2 on the right.
    The lines are sine fits to the data.}
    \label{fig:Figure_2_comparison_low_high_j}
\end{figure}

To quantify and compare the IL-DMI for several current densities
we assume the IL-DMI to act as a bias field that shifts
the hysteresis loop, as we have seen in \cref{fig:Figure_1_sample}b).
This effect can be quantified by an IL-DMI field,
which we name $H_{\rm{IL-DMI}}$. We define $H_{\rm{IL-DMI}}$
as the difference between the U-D and
D-U switching fields divided by two, similar to~\cite{avci_chiral_2021}.
\begin{equation}
    H_{\rm{IL-DMI}} = \frac{H_{\rm{SW,U-D}} + H_{\rm{SW,D-U}}}{2}
    \label{eq:H_DMI}
\end{equation}
with $H_{SW,U-D}$ and $H_{SW,D-U}$ the switching fields
for U-D and D-U switching, respectively, as illustrated in \cref{fig:Figure_1_sample}b).
This description assumes that there are no other bias effects leading to asymmetric
hysteresis loops, such as exchange
bias~\cite{stamps_mechanisms_2000, meiklejohn_new_1957}
or biquadratic interlayer exchange interaction effects~\cite{slonczewski_origin_1993}.
\Cref{fig:Figure_2_comparison_low_high_j} presents the values
of the $H_{\rm{IL-DMI}}$ field as function of $\phi$ for the two different
ferromagnetic layers for the lowest and highest current densities.
The sinusoidal dependence on the angle is the fingerprint
for the presence of an IL-DMI,
as the latter is directly dependent on the angle between the applied field
and the in-plane inversion symmetry breaking, i.e., the antisymmetric axis.
We can exclude any effect from biquadratic interlayer exchange,
which would be isotropic in nature~\cite{slonczewski_fluctuation_1991},
same with the DMI, which is symmetric in the field direction, i.e.,
$H_{\rm{SW}}(-\phi) = H_{\rm{SW}}(\phi)$~\cite{han_asymmetric_2016}.
Therefore, we can conclude that the IL-DMI is the origin of
the shift of the hysteresis curves at low and high current densities.

\begin{figure}[htb!]
    \centering
    \includegraphics[width=0.95\linewidth]{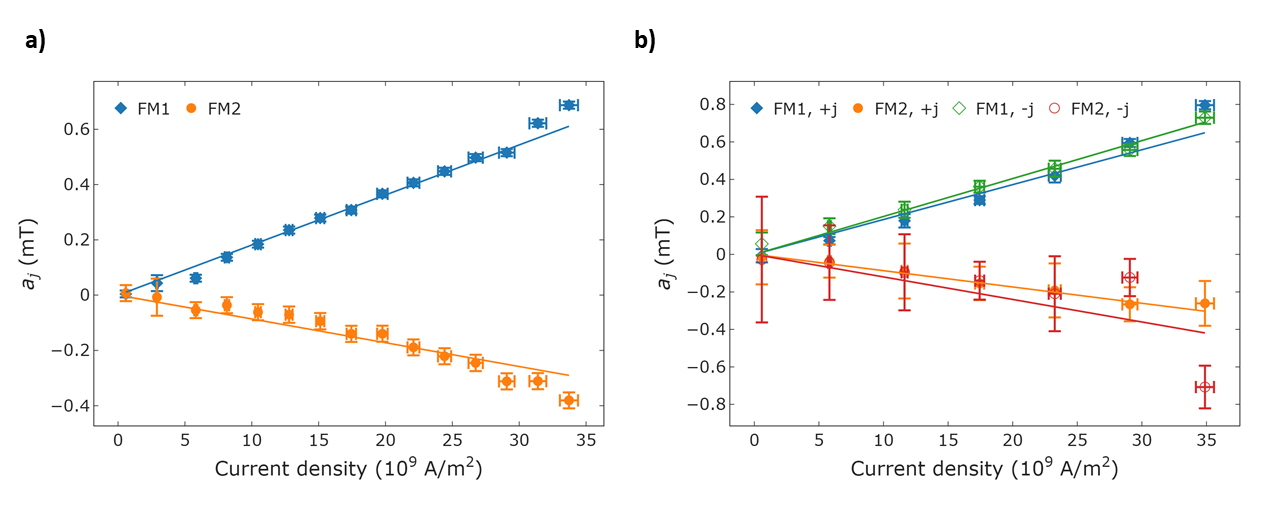}
    \caption{The current-induced contribution of the interlayer DMI $a_j$ as function of current density. a), b) display data from two different measurement sets. The first set in a) displays the strict current dependence for both ferromagnetic layers. In the second set the current polarity dependence was investigated. The displayed data $a_j$ is the amplitude of the current-induced part in the superimposed cosine fit as by \cref{eq:H_IL-DMI_model}. The lines represent linear fits to data up to $30\times10^9$A/m$^2$.}
    \label{fig:Figure_3}
\end{figure}

Next, we study the current-dependence of the $H_{\rm{IL-DMI}}$ for FM1 and FM2 as shown in \cref{fig:Figure_2_comparison_low_high_j}. 
There is a slight increase in the amplitude of $H_{\rm{IL-DMI}}$,
as well as a shift in the maximum position of the fitted sine function. 
The shift in the asymmetric axis indicates a change in the in-plane inversion
symmetry breaking. The latter is usually related to an asymmetric growth
or thickness gradient~\cite{han_long-range_2019}.
The application of a current should intrinsically break
the in-plane inversion symmetry as the electrons move in a defined
direction within the Hall-bar. Meaning when the current effect is significantly large it could change the direction of the in-plane symmetry breaking. Therefore, we can assume
that any applied electrical current yields an additional term to the
IL-DMI that is oriented along the direction of electrical current 
and can be described by a cosine function. To investigate this statement we devised
a simple model of two superimposed cosine functions, one static,
governed by the geometry of the sample and its intrinsic inversion-symmetry
breaking, and a second one for the current-induced effect.
We thus use the following model for a shared-parameter fit.
\begin{equation}
    H_{\rm{IL-DMI}} = a_0 \cos(\phi - \phi_0) + a_j \cos(\phi - \phi_j) + b,
    \label{eq:H_IL-DMI_model}
\end{equation}
with $\phi$ the angle of the applied field, $\phi_0$ the angular direction of symmetry breaking
defined by the sample growth, i.e., the maximum position of $H_{\rm{IL-DMI}}$
at zero applied current, $\phi_j$ the azimuthal direction of
the current-induced IL-DMI, $a_0$ and $a_j$ are the amplitudes
and $b$ is an offset. Note that the positive current direction
is along 0° (see \cref{fig:Figure_1_sample}).
We assume the first term to be independent of current,
thereby, $\phi_0$ is fixed along the direction of the antisymmetric axis 
and is determined by extrapolating sine fits of the data to the zero current limit which yields
$241\pm 3^{\circ}$ for the case of FM1 and $243 \pm 4^{\circ}$ for FM2. 
Furthermore, we assume $\phi_j$
to align with the current direction, i.e., towards 0° for positive current.
Using \cref{eq:H_IL-DMI_model} the data of $H_{\rm{IL-DMI}}$
for all available current densities is fitted.
We expect the parameters $a_0$ and $\phi_j$ to be constant for all
current densities, as long as the direction of the current does not change.
Therefore, we employ a shared-parameter fit, where $\phi_j \; \&\; a_0$
are shared parameters. We fix the value of $\phi_0$ to obtain $a_j$, which is not
a shared parameter as we expect it to increase with current density.
The resulting values of $\phi_j \; \&\; a_0$ are shown in 
\cref{tab:fit_parameters_current_dep} and show $\phi_j$ to point
approximately along 0°, i.e., the current direction.
The trend of $a_j$ with current density shows a linear increase,
see \cref{fig:Figure_3}a), however, the slope is positive for FM1
and negative for FM2, see slope in \cref{tab:fit_parameters_current_dep}. A slight increase for the slope at higher
current densities might be related to an increase in temperature
due to Joule heating.

\begin{table}[htb!]
    \centering
    \begin{tabular}{llrrr}\toprule
        current   & layer  & $\phi_j$ (°) & $a_0$ (mT)    & slope (µT/$10^9$Am$^{-2}$)  \\\midrule
        $+$ & FM1  & $-4.9 \pm 0.8$      & $0.48 \pm 0.01$ &  $18.1 \pm 0.3$\\
        $+$ & FM2  & $2.9 \pm 2.3$      & $0.48 \pm 0.01$ & $-8.6 \pm 0.3$\\\bottomrule
    \end{tabular}
    \caption{Shared parameters across all positive current densities
    obtained from fitting the IL-DMI field to current densities. And slope of the linear fits shown in \cref{fig:Figure_3}a).
    The initial value for $\phi_j$ is 0°.}
    \label{tab:fit_parameters_current_dep}
\end{table}

For better understanding of any possibly underlying asymmetry for FM1 and FM2,
the measurement was repeated using positive and negative current polarities.
If the proposed model is correct, then, due to reversal
of the current direction, $\phi_j$ should align towards 180°
while $a_j$ stays of similar magnitude. And indeed,
the shared parameter $\phi_j$ follows an alignment towards 180°
for negative current polarity, compare with 
\cref{tab:fit_parameters_current_polarity}. And $a_j$ displays similar
size for positive and negative fields within
the investigated current density range,
as shown in \cref{fig:Figure_3}b). The slopes of the linear fits displayed in \cref{fig:Figure_3}b)
are presented in \cref{tab:fit_parameters_current_polarity}. 
The nature of this observation can be better understood when employing symmetry arguments.
We may write the IL-DMI energy associated with the chiral state
as $F=(\textbf{T}_2-\textbf{T}_1)\cdot (\textbf{M}_1\times \textbf{M}_2)$, where
$\textbf{M}_1$ and $\textbf{M}_2$ are the magnetizations in FM1 and FM2, respectively,
and $\textbf{T}_i$ is the contribution to the torque on $M_i$ that stems from the IL-DMI (the total
torque is zero in the stationary state).
These torques $\textbf{T}_i$ need to point into the direction of $\textbf{M}_1\times \textbf{M}_2$ in order
to produce a contribution to $F$ (with $\textbf{T}_1\ne \textbf{T}_2$).
If the system contains a mirror plane that flips the sign of $\textbf{M}_1\times \textbf{M}_2$, it 
therefore also flips the relevant component of $(\textbf{T}_2-\textbf{T}_1)$ without
changing the energy $F$, which is inconsistent
with the DMI interaction that changes sign, when $\textbf{M}_1\times \textbf{M}_2$ changes sign.
This mirror plane may be eliminated by a material gradient or an applied
electric current in the direction
$(\textbf{M}_1\times \textbf{M}_2)\times \textbf{R}$, where $\textbf{R}$ is the distance vector
between the two ferromagnetic vectors. 
The injection of spin current from the SHE into the ferromagnetic layers
may generate these torques $\textbf{T}_2$ and $\textbf{T}_1$.
In this case one may understand the IL-DMI also from the balance equation of the
torques and it is not necessary to consider the energy $F$:
The torque balance equations on FM1 and FM2 may be written as
$-J(\textbf{M}_1 \times \textbf{M}_2) +\textbf{T}_1 =0$
and
$-J(\textbf{M}_2 \times \textbf{M}_1) +\textbf{T}_2 =0$,
where $J$ is the exchange coupling.
By summing these two equations one obtains
a single one:
$2J(\textbf{M}_1 \times \textbf{M}_2)=\textbf{T}_1-\textbf{T}_2$. Consequently, a chiral magnetic state
characterized by the chirality
$(\textbf{M}_1 \times \textbf{M}_2)\ne 0$ will occur when $\textbf{T}_2\ne \textbf{T}_1$. In the samples considered it is
likely that injection of spin current into the FM layers
yields even different signs of $\textbf{T}_2$ and $\textbf{T}_1$.

\begin{table}[htb!]
    \centering
    \begin{tabular}{llrrr}\toprule
        current   & layer  & $\phi_j$ (°) & $a_0$ (mT)    & slope (µT/$10^9$Am$^{-2}$)  \\\midrule
        $+$ & FM1  & $1.7     \pm 0.4$      & $0.43 \pm 0.02$       & $18.6 \pm 0.9$  \\
        $+$ & FM2  & $171.7   \pm 2.8$      & $0.44 \pm 0.04$       & $ -8.7 \pm 2.9$   \\
        $-$ & FM1  & $-5    \pm 20$         & $0.40 \pm 0.09$       &  $-20.2 \pm 0.6$ \\
        $-$ & FM2  & $222   \pm 13$         & $0.62 \pm 0.08$       &  $12.0 \pm 2.8$ \\\bottomrule
    \end{tabular}
    \caption{Shared parameters across all current densities obtained from fitting the IL-DMI field to positive and negative current densities from the additional measurement. The initial values for $\phi_j$ are -10° and 170°, for positive and negative currents respectively.}
    \label{tab:fit_parameters_current_polarity}
\end{table}

\begin{figure}[htb!]
    \centering
    \includegraphics[width=0.95\linewidth]{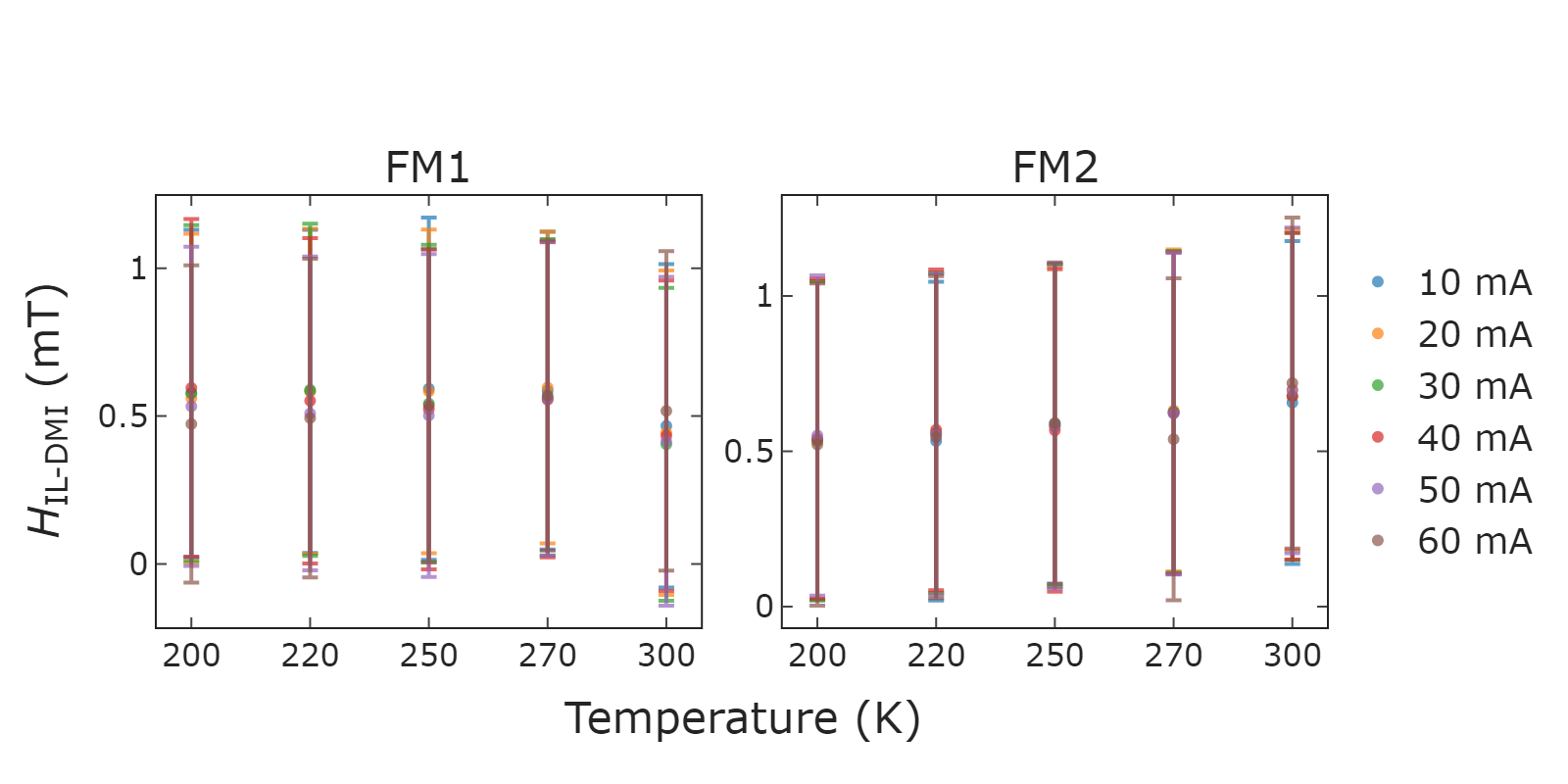}
    \caption{The IL-DMI field as function of temperature for a continuous film of Ta(4)/Pt(4)/Co(0.6)/Pt(0.5)/Ru(0.8)/Pt(0.5)/Co(1)/Pt(4) measured at various total currents.}
    \label{fig:Fig_4_cont_film_T_dep}
\end{figure}

The change of the effect due to the current poalrity shows that the effect does not arise due to heating effects.
To understand any effect heating may have on the sample,
a continuous film of a similar sample system
Ta(4)/Pt(4)/Co(0.6)/Pt(0.5)/Ru(0.8)/Pt(0.5)/Co(1)/Pt(4),
was investigated in a cryostat as function of temperature. 
The effect on the IL-DMI field for temperatures
varying between 200 K and 300 K is present but small, see 
\cref{fig:Fig_4_cont_film_T_dep}. 
Additionally, the Joule heating in the Hall bar was simulated using COMSOL multiphysics, similar to an 
approach to estimate the heating in NiO/Pt systems~\cite{meer_direct_2021}. 
The simulation yields a Joule heating of about 18 K for a current density of $38.4\,\times 10^9 $A/m$^2$.
Considering the small dependence of IL-DMI on temperature, as seen in \cref{fig:Fig_4_cont_film_T_dep}, 
and the little effect of Joule heating, we can deduce Joule heating to play a negligible effect on the magnitude of the IL-DMI in our sample.  

\section{Summary}

In conclusion, the current dependence of the IL-DMI in SAFs
has been found by studying the switching fields obtained from
AHE measurements and quantified by means of the resulting
IL-DMI field $H_{\rm{IL-DMI}}$, which displays a significant shift
in the azimuthal dependence for increasing current density.
By employing a simple model of two superimposed cosine functions
with a static and current-induced contribution, the observed behavior
was reliably modeled. The fit yields a current-induced
contribution for the IL-DMI, scaling linearly with current. The opposite current polarity reveals
a consistent trend, with the opposite azimuthal dependence and similar amplitudes.
From the symmetry point of view, an applied electric current along the direction $(\textbf{M}_1\times \textbf{M}_2)\times \textbf{R}$ --
where $\textbf{R}$ is the distance vector between the magnetizations $\textbf{M}_1$
and $\textbf{M}_2$ of the
two layers -- allows for a chiral Néel-type
state.
One possible mechanism for the IL-DMI is the spin Hall effect,
which provides torques of opposite sign on the two magnetizations FM1 and FM2.
The observed effect is conforming with the performed symmetry analysis, 
calling for future studies to also develop a quantitative theoretical model.
The effect has the appearance of a purely current-induced effect, as growth dependent IL-DMI and current-induced IL-DMI superimpose.
Therefore, we assume this effect to be present in systems without IL-DMI.
Such a current-induced contribution to the IL-DMI allows for the highly desirable
control of this chiral interaction and enables a direct,
post-growth tuning mechanism for 3D topological spin structures
in future spintronic devices.

\section{Acknowledgements}
F.K. acknowledges funding by the Deutsche Forschungsgemeinschaft (DFG, German Research Foundation) - TRR 173/2 - 268565370 Spin+X (Projects A01+B02).
This project has received funding from the European Research Council (ERC) under the European Union's Horizon 2020 research and innovation programme (Grant No. 856538, project “3D MAGiC”).

\printbibliography

\end{document}